# Anomalous magnetic, transport and thermal properties of Gd$_5$Ge$_3$

Bibekananda Maji[1], K. G. Suresh[1,*] and A. K. Nigam[2]

[1]Department of Physics,

Indian Institute of Technology Bombay, Mumbai- 400076, India

[2]Tata Institute of Fundamental Research,

Homi Bhabha Road, Mumbai- 400005, India

*Abstract*

We have studied the magnetic, thermal and magnetotransport properties of polycrystalline Gd$_5$Ge$_3$. It is found that the compound is a strong antiferromagnet and does not undergo any strong metamagnetic transition even in a field of 90 kOe. However, a small but visible ferromagnetic component is found to coexist with the antiferromagnetic order at low temperatures, as revealed by the anomalies in magnetization, magnetoresistance and heat capacity data. Our data suggest that the ferromagnetic component is of magnetostructural in origin, as evidenced by the strong field induced irreversibility of the magnetoresistance and heat capacity isotherms. The 'first order like' magnetostructural distortion established in this material is found to result in a martensitic like scenario wherein the kinetic arrest controls the field/temperature dependence of physical properties such as heat capacity and magnetoresistance.

*Corresponding author (email: suresh@phy.iitb.ac.in, Fax: +91-22-25723480)

**I. Introduction**

Intermetallic compounds formed between rare earths and nonmagnetic elements have become very attractive because of various anomalous properties exhibited by them. One of the reasons for the interest in these materials is that some of them undergo field induced magneto-structural transition, giving rise to many remarkable phenomena such as large magnetoresistance, colossal magnetostriction, giant magnetocaloric effect etc. The anomalous electronic, magnetic and thermal properties exhibited by many of these compounds have been intensively investigated over more than a decade.[1-5] The martensitic-like nature of the transition and the strains resulting from the structural change or even a field induced lattice distortion is found to influence the magnetic and the related properties in some of these materials. Rare earth germanides and silicides are good candidates of this family.[1,3,5-7] Recently, a field-induced metamagnetic-like transition was reported in single crystals of $Gd_5Ge_3$ at a critical field ($H_c$) of ~ 38 kOe along the *a* axis.[8] The authors have attributed this to a field induced lattice deformation. Later, the irreversible nature of the lattice deformation was also studied by magnetostriction and thermal expansion measurements.[9]

It is known that $Gd_5Ge_3$ crystallizes in the $Mn_5Si_3$ type hexagonal crystal structure with the space group *P6₃/mcm*. The Gd atoms occupy two crystallographically inequivalent sites, namely *6(g)* and *4(d)*, whereas Ge atoms occupy *6(g)* sites. It has a repeated layer of *Gd-4(d)* atoms and *(Gd+Ge)-6(g)* atoms along the *c-* axis. It has been reported that this compound exhibits an antiferromagnetic transition at 76 K (Neel temperature, $T_N$) and another one at 52 K ($T_t$).[10] In a recent report, it is confirmed form the temperature variation of x-ray powder diffraction experiment that $Gd_5Ge_3$ undergoes a structural distortion from hexagonal to orthorhombic structure at its $T_N$.[11] The authors have also mentioned about the weakly first order nature of the antiferromagnetic transition, as there is a small volume change at the transition temperature. They have further shown that the structural distortion in the sample prepared with commercial Gd is weak. The detailed magnetic structure has not yet been established experimentally for this compound. However, a model of the magnetic structure is proposed by Narumi et al.

in which Gd atoms on the 6g sites form a trimer, whose spins form an antiferromagnetic triangular lattice in the *a-b* plane and are aligned parallel to each other along the *c*-axis. The Gd spins on the 4(d) site order antiferromagnetically in the *a-b* plane and are coupled ferromagnetically along the *c*-axis.[8]

Recently, we have reported the magnetic properties of $Nd_5Ge_3$, which is the analogue of $Gd_5Ge_3$[12]. Spontaneous magnetization jumps and field induced irreversibility in the magnetization, heat capacity and magnetoresistance isotherms at low temperatures were observed in this compound. However, there have not been many attempts to explore the detailed magnetic and related properties of $Gd_5Ge_3$. Though both these compounds belong to the same series, the absence of strong crystal field effect in $Gd_5Ge_3$ makes the magnetic and related properties of this compound somewhat different from that of $Nd_5Ge_3$. Therefore, in this paper, we report the results of our investigations on the magnetic, transport and thermal properties of polycrystalline $Gd_5Ge_3$.

**II. Experimental details**

Polycrystalline sample of $Gd_5Ge_3$ was prepared by arc melting stoichiometric mixture of the constituent elements of Gd (99.9 - at. % purity) and Ge (99.999-at. % purity) in a water-cooled copper hearth, in high purity argon atmosphere. The resulting ingot was turned upside down and remelted five times to ensure homogeneity. The weight loss after the final melting was less than 0.5 %. The arc melted button was sealed in an evacuated quartz tube and annealed at 1000 °C for 1 week. The structural analysis of the annealed sample was performed by collecting the room temperature powder x-ray diffractogram (XRD) using Cu-K*α* radiation. The magnetization measurements were carried out using a vibrating sample magnetometer attached to a Physical Property Measurement System (Quantum Design, PPMS-6500). The heat capacity ($C_p$) and the electrical resistivity (ρ) measurements were also carried out in PPMS.

**III. Results**

Rietveld refinement of the XRD pattern of $Gd_5Ge_3$ at room temperature confirms that there are no detectable impurities in the compound. The crystal structure is found to be hexagonal $Mn_5Si_3$- type with the space group $P6_3/mcm$ (193). Lattice parameters calculated from the refinement are $a = b = 8.573$ Å and $c = 6.461$ Å. Fig. 1a and b shows the temperature dependence of dc magnetic susceptibility ($\chi$) and heat capacity in different fields. From the $\chi$-T data, it is clear that the compound undergoes an antiferromagnetic transition at $T_N = 81$ K. It is to be mentioned here that different $T_N$ values have been reported by different authors for this compound.[10-15] Recently, Mudryk *et al.* have reported the highest $T_N$ value of 87 K along with the observation that the $T_N$ in this compound is strongly affected by the purity of Gd.[11] Below $T_N$, the magnetization decreases due the antiferromagnetic order and again shows an up-turn giving rise to a cusp at about 52 K, in agreement with the reported transition at $T_t$.[10] Both these transitions are also shown by zero field heat capacity, as shown in Fig. 1b. Khushwaha and Rawat very recently have reported a $T_N$ of 82 K and a $T_t$ of 50 K[16]. These authors have also reported a third transition at 36 K. As can be seen from Fig. 1, with the application of field, both the peak positions (at $T_N$ and $T_t$) shift towards low temperature. This indicates that both the transitions are of antiferromagnetic in nature. The field dependence of the heat capacity peak height that is seen in the present case is almost identical to that observed by Mudryk et al.[12], indicating a reasonably 'first order-like' transition at $T_N$ in the present case too, though we have used commercial Gd.

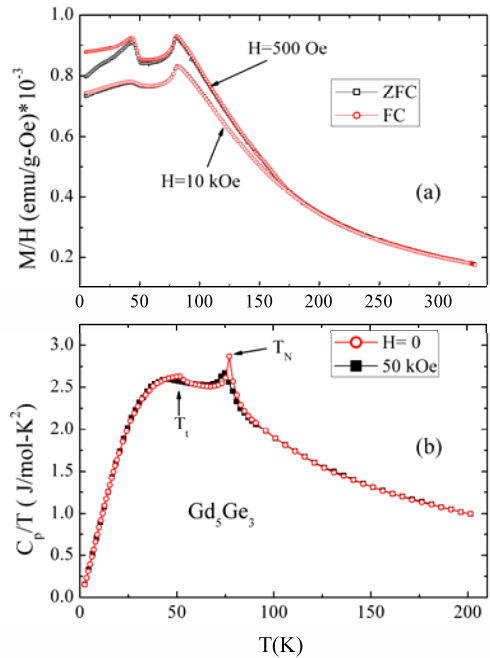

FIG. 1. Temperature dependence of dc magnetic susceptibility and heat capacity in different fields.

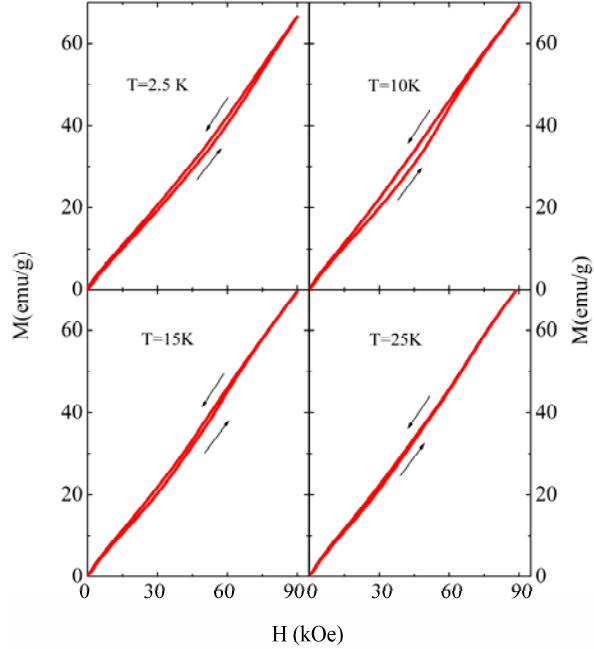

FIG. 2. Isothermal M *vs.* H curves of $Gd_5Ge_3$ at 2.5 K, 10 K, 15 K and 25 K.

Fig. 2 shows the M(H) curves recorded at some selected temperatures below $T_N$. There is no trend of saturation even up to 90 kOe, which indicates the strong antiferromagnetic order present in the compound. The remanent magnetization is zero at all the temperatures. However, Narumi et al.[8] have reported the magnetization isotherm in pulsed magnetic field up to ~ 600 kOe in $Gd_5Ge_3$ single crystal and showed that magnetization saturates above 300 kOe for both a- and c- axes. Interestingly, in the present case, at 2.5 K, the M(H) curve exhibits a small but noticeable hysteresis. The opening of the M-H loop is found to be maximum at 10 K and at 25 K, the curve is almost anhysteretic. This is probably indicative of the development of a small ferromagnetic (FM) component at low temperatures. It may also be noted that the slope of the M-H loops shows some changes indicative of some weak metamagnetic like transitions. Therefore, the low temperature phase appears to have the coexistence of a weak ferromagnetic and a strong antiferromagnetic components.

Temperature dependence of normalized resistivity (inset of Fig. 3) curves measured from 5 to 300 K in different fields exhibit a change in slope at ~77 K and ~50 K corresponding to $T_N$ and $T_t$, respectively, as found in the magnetization and heat capacity data. More importantly, the resistivity curves at 50 kOe and 80 kOe show a minimum at low temperatures (~ 10 and ~15 K respectively). It is to be mentioned here that $Gd_5Ge_3$ is known to possess large spontaneous magnetostriction at low temperatures due to strong magnetoelastic coupling.[8,9,16] The magnetostriction is further enhanced by the application of the field.

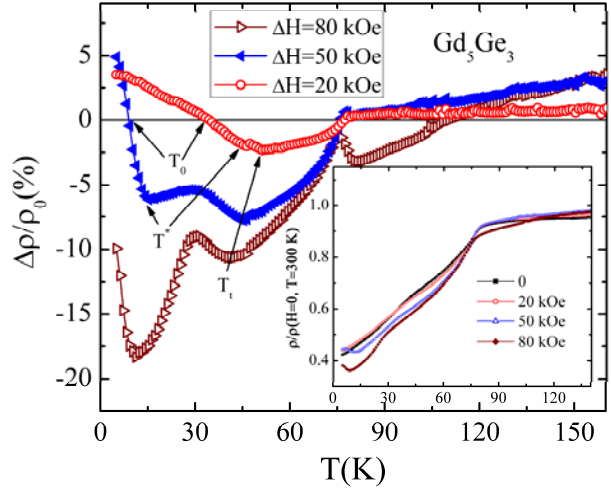

FIG. 3. Temperature dependence of magnetoresistance of $Gd_5Ge_3$ in different fields. Inset shows the normalized resistivity vs. temperature in different fields.

The resistivity in the higher temperature region increases slightly with field. This leads to positive magnetoresistance, MR (defined as $MR(\%) = \left( \frac{\rho(H) - \rho(H=0)}{\rho(H=0)} \right) \times 100$) in the higher temperature region which is rather unexpected [Fig. 3]. A similar observation of positive MR in the paramagnetic regime was also observed in $Gd_5Ge_4$ and itinerant electron ferromagnet $LaFe_4As_{12}$.[17,18] In the main panel of Fig. 3, we show the variation of MR as a function of temperature from 5 to 160 K for different fields. At 20 kOe, the MR is low and positive throughout the temperature range investigated, except for a small regime between $T_N$ and $T_t$ (seen in the $M$-$T$ or $C_P$-$T$ data). The positive MR can be attributed to the Lorentz force effect as observed in other metallic systems.[18] It can also arise due to anisotropic magnetoresistance or domain rearrangement. The negative MR seen between $T_N$ and $T_t$ suggests the presence of a ferromagnetic component in this temperature interval.

For fields of 50 and 80 kOe also, the overall trend remains the same as at 20 kOe. However, the temperature regime over which the MR is negative is found to be larger at 50 and 80 kOe. There is a dip near $T_t$ which is found to shift towards low temperatures with increase in field. The tendency of MR to become positive at low temperatures, causing a minimum at $T^*$ in all fields indicates the onset of a strong positive contribution. It can be assumed that the negative contribution from the ferromagnetic component is frozen around $T^*$ and as a result, the positive contribution from conduction electrons dominates below this temperature.[19] Another possibility is that the applied field induces gaps in the Fermi surface, leading to a reduction in the effective number of conduction electrons. This will reflect as an increase in the resistivity with decreasing temperature in the antiferromagnetic state. One can notice that the minimum at $T^*$ becomes very prominent as the field increases and also it shifts towards low temperatures. The minimum is found to appear at about 15 K and 11 K in 50 and 80 kOe respectively. The competition between positive and negative contributions gives rise to zero MR at $T_0$ where both the contributions are nearly equal in magnitude. The positive contribution is found to dominate below $T_0$, giving rise to positive magnetoresistance. At 80 kOe, though the minimum is clearly seen, the MR at the lowest temperature is still negative. This could be due to the strong enhancement of the FM component (as evidenced by the opening of the hysteresis loops shown in Fig. 2). It is also possible that a structural contribution is also present which causes negative MR. It may be recalled here that the magnetic hysteresis loops showed the maximum opening at about 10 K. The minimum MR values are found to be -6 % and ~ -18 % at around 15 K and 11 K for H= 50 kOe and 80 kOe, respectively. By comparing the magnetization and the MR data, it is evident that the sharp change in the sign and the magnitude of MR with temperature is in tune with the development of the FM component mentioned above. The coexistence of the two components gives rise to anomalous MR variation. Therefore, it is seen that the MR variation at low temperatures is determined by the competing effects of antiferromagnetic order and the weak ferromagnetic component. The existence of the competing magnetic phases mentioned above is further probed with the help of MR and heat capacity isotherms, as shown below.

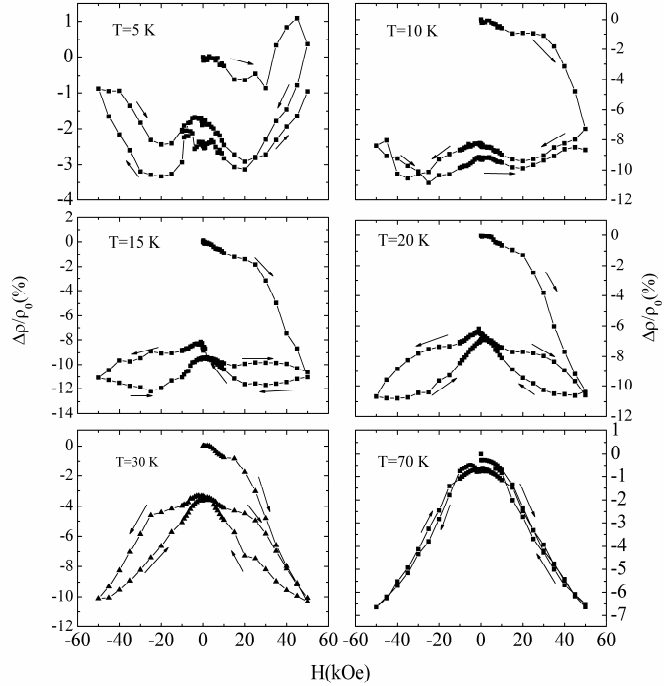

FIG. 4. Field dependence of magnetoresistance for $Gd_5Ge_3$ at a few selected temperatures

Isothermal MR curves at various temperatures are shown in Fig. 4. Before recording the data as a function of field, the measurement temperature was reached by cooling the sample in zero magnetic field from the paramagnetic region (300 K) every time. For T=5 K, the virgin curve exhibits nominally positive MR in the field range of 30 to 50 kOe. However, in all other curves and for other temperatures, the MR is negative throughout. Therefore, it is quite evident that the positive contribution is dominant only at 5 K. It is important to note that the (negative) MR in zero field after field cycling is about 9 % higher than the respective virgin zero field values at 10, 15 and 20 K. Therefore, there is a field-induced irreversibility in the MR value. Furthermore, the MR isotherms display butterfly-like open hysteresis loops at almost all the temperatures, but prominently in the region 10 K ≤ T ≤ 30 K, but almost disappears at 70 K. The irreversibility in MR is generally observed in systems where field induced magnetic and structural transitions take place. In such cases, a fraction of high field magnetic phase gets supercooled and kinetically arrested during field cycling. Almost identical observations have been obtained[19] in the case of Si doped $CeFe_2$. Si doping in this case stabilizes the antiferromagnetic ground state. Application of a field partially converts the antiferromagnetic phase into a ferromagnetic phase. This magnetic transition is accompanied by a structural change. The coexistence of these two structurally and magnetically different phases gives rise to interesting magnetic, MR and heat capacity properties. Therefore, a similar scenario is likely to be present in the case of $Gd_5Ge_3$ as well. The fact that the virgin curve lies outside the envelope curve in these MR isotherms is another indication about the presence of first order magnetostructural effects.[20] Therefore, the data seen in Fig. 4 show that there are competing antiferromagnetic and ferromagnetic phases in the temperature range of 5-30 K, which decreases at higher temperatures.

It is important to mention here that, quite contrary to the present MR data, the recent report by Kushwaha and Rawat[16] has shown large positive MR in the temperature range of 5- 60 K and a moderate negative MR at 100 K for a maximum field of 80 kOe. Interestingly, they have also observed the strong field induced irreversibility in the MR isotherms. MR was found to become increasingly positive after the virgin cycle of magnetization.

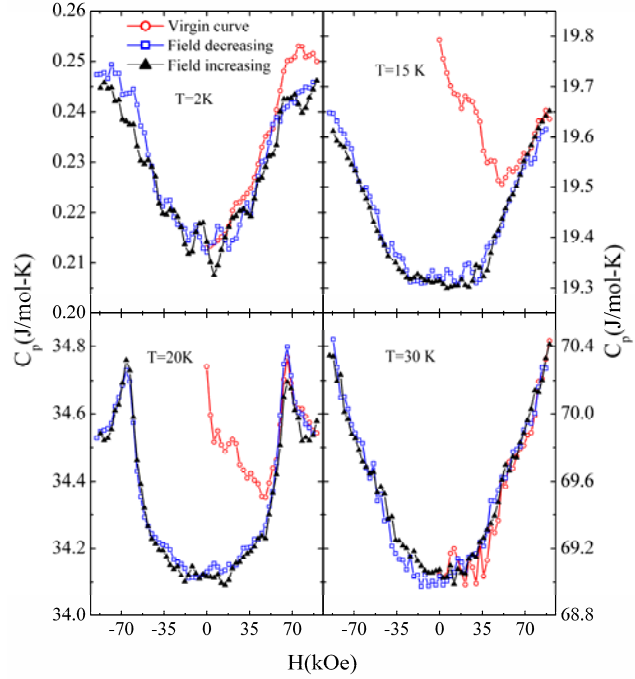

FIG. 5. Heat capacity of $Gd_5Ge_3$ as a function of magnetic field at 2 K, 15 K, 20 K and 30 K

In order to shed more light on the irreversibility, we have measured the isothermal heat capacity as a function of *H* from -90 kOe to +90 kOe at 2, 15, 20 and 30 K after cooling the sample in zero field from 300 K each time (Fig. 5). For T=2 K, $C_p$ increases with H and it comes back to the zero field value after the removal of the field. The curve is symmetric for the positive and negative field axes. As the temperature is increased to 15 K, the zero field heat capacity value increased considerably. At this temperature, on increasing field, there is a sharp fall in the heat capacity up to about 50 kOe. Further rise in the field leads the heat capacity to increase. On returning from 90 kOe the $C_p$ vs. H curve separates from the virgin curve roughly below 65 kOe and decreases continuously. The zero field $C_p$ values after field cycling remain lower than the virgin zero field value. Almost a similar behavior is observed at T= 20 K which also shows a difference between virgin $C_p$ (H=0) and remnant $C_p$ (H=0) values. However, unlike at 15 K, there is an additional feature in the form of a sudden down turn at about 66 kOe, both in the positive and negative sides. The behavior of the $C_P$ isotherms at 15 and 20 K clearly indicates the presence of competing magnetic or magnetostructural phases. It is interesting to note that the field induced irreversibility of the heat capacity behavior is closely related to the opening of the hysteresis loop (shown in Fig. 2). Furthermore, the strong irreversibility under field reversal supports the presumption of the magnetostructural contribution accompanying the FM component. It is also of importance that overall, the behavior of these isotherms is identical to that of the MR isotherms. A similar reversal of the trends of the $C_p$-H plots has been observed in Si doped $CeFe_2$, which is known to have a coexistence of competing ferro and antiferromagnetic components.[21] Irreversibility in the $C_p$-H curve is reported in (Tm,Tb)$Co_2$ and (Er,Y)$Co_2$, which undergo a field induced structural distortion at their Curie temperatures.[22, 23]

## 1V. Discussion

We find that though there are some reports on magnetic and related properties of $Gd_5Ge_3$ in the literature, the underlying magnetic state of this compound remains not fully understood yet. It is quite clear that the material retains its antiferromagnetic order even at the highest field of 90 kOe. The major observation of this study is the strong field induced irreversibility seen in the heat capacity and the MR isotherms at temperatures above 2.5 K. Another important observation is the opening up of the M-H hysteresis loop, again in the same temperature interval. These observations are characteristic of (i) the coexistence of some weak ferromagnetic phase along with the predominant antiferromagnetic phase and (ii) magnetostructural coupling. Mudryk etl al.[11] have attributed the evolution of the ferromagnetic phase to spin reorientation transition. Based on the present data, we feel that the magnetostructural distortion gives rise to a FM component at temperatures below $T_N$. The strong field induced irreversibility suggests the presence of 'first order like' effects, which are of magnetostructural in origin. This is even more significant in the context of the recent detailed field and temperature variation of XRD study by Mudryk et al.[11] Interestingly, magnetostriction study on the same system has also shown a similar field induced anomaly at low temperatures.[10,16] The magnetostriction at 5 K is of irreversible character, i.e., the distortion caused by the field is stored and the sample length does not come back to the original value after ramping the field to zero. The thermal expansion curves of zero field cooling and the heating after the application of field do not follow the same path at low temperatures (below 20 K).[9] All these observations suggest that the material undergoes lattice distortion due to the magnetic field, which can influence heat capacity and resistivity. The magneto-structural distortion gives rise to the weak FM component, even while the dominant magnetic order is antiferromagnetic. Application of a field causes a weak metamagnetic-like transition, as reflected by the slope changes in the M-H curves (Fig.2). The coexistence of the FM and AFM phases gives rise to the interesting transport and thermal properties. In summary, we feel that the most important aspect of the present work is that it could demonstrate the role of magnetostructural distortion in determining various properties, though the distortion is reported to be quite weak in samples prepared with commercial Gd. It is felt that a probe like $^{155}$Gd Mössbauer spectroscopy will be essential to confirm/rule out the spin reorientation transition proposed in this material. Once that is done, a clear picture on the origin of the now established weak FM component would emerge. A complete understanding of the magnetism in this compound will be of great importance in the research on several such materials of practical importance.

## V. Conclusion

In this work, we have studied the magnetic, thermal and transport properties of polycrystalline $Gd_5Ge_3$ prepared using commercial Gd. Basic crystallographic and magnetization data are in overall agreement with most of the reported studies. It is found that the compound is a strong antiferromagnet and does not undergo any strong metamagnetic transition even in a field as large as 90 kOe. However, a small but visible ferromagnetic component is found to coexist with the antiferromagnetic order at temperatures below the Neel temperature. The ferromagnetic component seems to be of magnetostructural in origin. Coexistence of competing magnetic and structural phases gives rise to complex MR and heat capacity isotherms. The 'first order like' magnetostructural distortion established in this material is found to result in a martensitic like scenario wherein the kinetic arrest controls the field/temperature dependence of physical properties such as heat capacity and magnetoresistance.

**Acknowledgement**

The authors thank D. Buddhikot for his help in the resistivity measurements.